\documentstyle[12pt,epsfig,epsf]{article}                                                   
\begin{document}                                                                

\title{\={p}p low energy parameters from annihilation cross section data}

\author{J. Carbonell, K.V. Protasov, \\
\small
Institut des Sciences Nucl\'eaires, 53, Avenue des Martyrs,
38026 Grenoble, Cedex, France, \\
\large
A. Zenoni, \\
\small
Dip. di Chimica e Fisica per l'Ingegneria e per i Materiali, \\ 
\small
Universit\`a di Brescia and INFN, Sez. di Pavia, Italy }
\date{ }
\begin{titlepage}
\maketitle
\thispagestyle{empty}

\begin{abstract}
\normalsize
{\bf Abstract.} 
The recent experimental data obtained by the OBELIX group 
on total $\bar{p}p$ annihilation cross section are analysed;
the low energy spin averaged parameters of the $\bar{p}p$ scattering amplitude 
(the imaginary parts of the S-wave 
scattering length and P-wave scattering volume) are extracted from the data.
Their values are found to be equal to
$\mbox {Im} a_{sc} = - 0.69 \pm 0.01 ($stat$) \pm 0.03 ($sys$) \mbox { fm},$
$\mbox {Im} A_{sc} = - 0.76 \pm 0.05 ($stat$) \pm 0.04 ($sys$) \mbox { fm}^3$.
The results are in very good agreement with existing atomic data. 
\end{abstract}
\vfill
Published in Physics Letters {\bf B 397} (1997) 345$-$349
\end{titlepage}

\section{Introduction}

Recent LEAR data allow one very interesting question 
(from the point of view of general quantum mechanics) to be investigated 
experimentally: the behaviour of the total reaction cross
section near threshold in a hadronic system with Coulomb attraction.

This is made possible by the measurement, performed for the first time
by the OBELIX experiment, of the
$\bar{p}p$ annihilation cross section at very low antiproton momentum, from
175 MeV/c down to 44 MeV/$c$ \cite{OBELIX,LEAP96}.
Starting from the article of Wigner \cite{Wigner} it is known that, in
such a system, the usual $1/v$ near threshold behaviour of the exothermic
reaction cross
section should be replaced by the $1/v^2$ one, $v$ being the velocity of the
incident particles. This fact is known in
quantum mecanics \cite{Baz,Landau,Newton} but, until these results were obtained
at LEAR, the question was of academic interest in nuclear physics.

Now, the cross section data from the OBELIX experiment make possible 
not only to see, for the first time, the
$1/v^2$ behaviour in such a hadronic system, but also to understand at
what energies this regime becomes evident. In fact, it appears that the Coulomb
forces, responsible for this change of behaviour, induce a very significant
modification of the annihilation cross section at antiproton momenta 
higher than the naive estimated value, which is of the order of the inverse 
Bohr radius of protonium.
Moreover, these data can be used to obtain information about the low
energy parameters of the $\bar{p}p$ system (S-wave scattering length
and P-wave scattering volume), as it was shown in \cite{CP93}. Up to now,
these data represent the only source, alternative and complementary
to atomic data \cite{CP92}, from which the values of these parameters can be 
extracted.

The aim of this article is to analyse the low energy $\bar{p}p$ cross section, 
to extract the imaginary parts of S-wave scattering length and P-wave 
scattering volume
and to compare the results with the ones obtained from atomic data.

\section{Antiproton-proton annihilation cross section}

At low energies, the total reaction (e.g. annihilation)
cross section for neutral particles
is well known to behave as $1/v$. If one looks for the next term in the 
development of the cross section in the center of mass momentum $q$, one
can see that the two first terms are entirely defined by the imaginary
part of the scattering amplitude $a_s$
\footnote{Hereafter we use the definition of scattering length
with negative imaginary part.} \cite{Shapiro}:
\begin{eqnarray*}
\sigma_{ann}^{s}={\pi\over q^2}\, \left(1 - |S|^2 \right) \approx
{\pi\over q^2}\, \left(1 - \mbox{e}^{2 {\rm Im} a_s q} \right) \approx
4 \pi \left({\mbox {Im} (-a_s) \over q} - 2\mbox {Im}^2 (-a_s) \right).
\end{eqnarray*}
This circumstance allows the imaginary part of the scattering
amplitude to be extracted from the behaviour of the total reaction 
(annihilation) cross
section, for instance from the antineutron-proton annihilation cross 
section \cite{Mutchler}.

An analogous expression can be obtained for an interacting system
with Coulomb interaction, for instance for the $\bar{p}p$ system. This problem
was solved in \cite{CP93}, where it was shown that the annihilation
cross section for the S-wave can be written as:
\begin{eqnarray}
q^2\sigma_{ann}^{sc} ({\mbox{S-wave}}) =  {8 \pi^2 \over 1-e^{2\pi\eta} }\,
\frac{{\mbox {Im}}(- a_{sc}/ B)}{|1+iq w(\eta) a_{sc} |^2}.
\end{eqnarray}
where:
\begin{itemize}
\item[$-$] $\eta=-1/qB$ is the dimensionless Coulomb parameter with
$B$ the $\bar{p}p$ Bohr radius;
\item[$-$] $w(x)=c_{0}^2(x)-2ix h(x)$ is an auxiliary function with
$qBw(\eta)\rightarrow2\pi$ when $q\rightarrow 0$;
\item[$-$] $c_{0}^2$ and $h$ are the usual functions in the Coulomb
scattering theory
\begin{eqnarray*}
c_{0}^2(x)  =  {2\pi x\over\exp(2\pi x)-1}; \hspace{1cm}
h(x)        = {1\over2}\left[\Psi(-i x)+\Psi(i x)\right]-
{1\over2}\ln\left(x^{2}\right)
\end{eqnarray*}
with the digamma function $\Psi$.
\end{itemize}
This expression is written within the usual scattering length approximation,
$a_{sc}$ being the scattering length of the strong interaction in presence
of Coulomb forces. 

Let us make few comments about this expression of the annihilation 
cross section.

First, this formula is written for spinless particles just to simplify the
expression and because of the impossibility to extract, from the present 
experimental data, triplet and singlet scattering lengths separately. 
Otherwise, it should be
necessary to write the total annihilation cross section as a sum of singlet
and triplet partial cross sections, with the usual statistical weights 
1/4 and 3/4 and as function of the corresponding spin dependent 
scattering lengths.
The reason why the two contributions cannot be determined separately
from the annihilation cross section data is connected to the fact that,
unless the values of the triplet and singlet scattering lengths are very 
different, they present essentially the same behaviour with energy.

Strictly speaking, the average value of the amplitude (Im $a_{sc}$) obtained
hereafter is not an usual average value of singlet and triplet amplitudes
($\frac{1}{4}$Im $a$(s) + $\frac{3}{4}$ Im $a$(t)), but some effective average value
obtained by averaging the cross sections.
The difference between these two values appears due to presence of the
scattering length in the denominator of (1).
One can easily estimate the error introduced by this approximation. Even if
the triplet scattering length is two times smaller than the singlet one the
difference between the two definitions of average value does not exceed two
percents.

Second, this approximation works, for the S-wave, within few percent accuracy 
up to antiproton laboratory momentum of 100 MeV/$c$ where, in principle, it 
becomes
necessary to account for the charge exchange cross section \cite{CP93}.

A third and rather general comment is aimed to remark the fact that, 
in the expression for the absorbtion cross section, not only the imaginary part
but also the real part of the scattering amplitude appears. 
Unfortunately, the contribution to the cross section due to the real part 
of the scattering length is very small; in fact it is of the order of 
few percents at the highest momentum considered.
Therefore, at the present level of experimental precision, the data can not be
used to extract the real parts too. Nevertheless, it is worth
emphasizing that the contribution to the cross section, due to the real part
of the scattering length, appears only thanks to the presence
of the Coulomb interaction. In some sense the situation recalls the case
of the Coulomb-nuclear interference in elastic scattering, where the sign of
the real part of the nuclear amplitude can be determined only thanks to 
the presence of the Coulomb forces.

If one wants to describe the experimental data up to 100 MeV/$c$ it
is necessary to take into account the P-wave contribution too, 
which is of course small, but not negligible. 
This generalization can be easily done by replacing
the scattering length $a_{sc}$ by
$A_{sc}(1+\eta^2)q^2$ where $A_{sc}$ is the corresponding P-wave
spin averaged scattering volume \cite{CP96}: 
\begin{eqnarray}
q^2\sigma_{ann}^{sc} ({\mbox{ P-wave}}) =  24 \pi^2 \, {1 + 1/\eta^2 
\over 1-e^{2\pi\eta} }\, \frac{{\mbox {Im}}(- A_{sc}/ B^3)}
{|1-q w(\eta)(1 + 1/\eta^2){\mbox {Im}}A_{sc}/B^3 |^2}.
\end{eqnarray}
The correction in the denominator, due to the real part of the scattering 
volume, can be omitted since, for P-wave, its contribution is negligible.

The remarkable property of the P-wave cross section
(and of other higher order partial waves) consists in the fact
that it does not vanish even at zero energy! Like S-wave reaction
cross section, the P-wave zero energy limit $q^2\sigma_{ann}^{sc}$ is equal
to a constant. Actually, the value of this constant is small
in comparison with the S-wave one, in fact the ratio of P-to-S contributions is
approximately equal to 3 Im$A_{sc}$/Im $a_{sc} B^2$, 
nevertheless it is not equal to zero. It means that if one measures the
annihilation into a particular channel for which
the S-wave annihilation is forbidden, nevertheless, the annihilation
cross section would be different from zero even at very low energy due to
P-wave contribution.

\section{Analysis of the experimental data}

\begin{figure}[h!]
\begin{center}
\epsfxsize=10cm
\epsfysize=15cm
\mbox{\epsffile{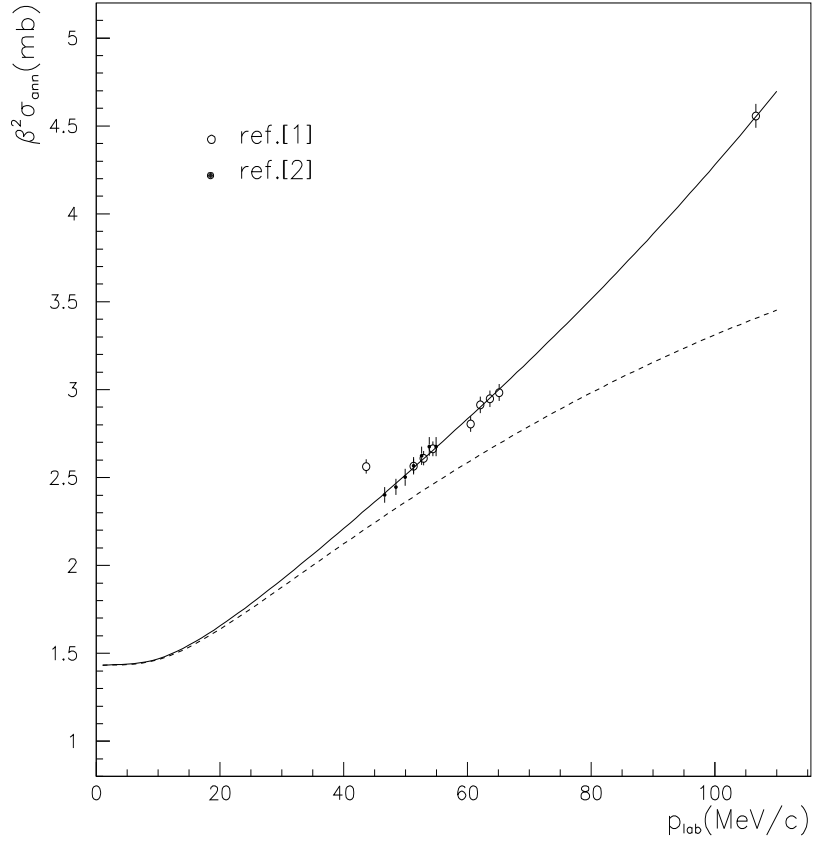}}
\end{center}
  \caption{Values of the total $\bar{p}p$ annihilation cross section 
           multiplied by the square of the incoming beam velocity. 
           Open dots are from Ref. [1], full dots are from Ref. [2]; the
           error bars represent the statistical error of the data.
           The point at 43.6 MeV/c is not included in the fitting procedure.
           The theoretical curves are the result of the present work:
           the full line is the total annihilation cross section, the dashed 
           line represents the S-wave contribution.
   }
  \label{F1}
\end{figure}

The data considered for the fitting of the theoretical expression of the
annihilation cross section are those obtained by the OBELIX experiment 
at LEAR \cite{OBELIX,LEAP96}. The data are reported in Tab.~\ref{TAB1}, where
the values of the cross section multiplied by the antiproton velocity 
$\beta\sigma^T_{ann}$ are given, together with the statistical and systematic
errors. 
Among these data the point at 174.4 MeV/c 
was not used for the fit since, at this value of the incident momentum, 
the scattering length approximation of the cross section is no longer valid. 
Moreover, as pointed out in ref.~\cite{LEAP96}, the
point at 43.6 MeV/c appears somehow in disagreement with the trend of the other
data, due to possible systematics out of control at this extremely low 
incident momentum value. 
So we prefer not to include it in the fitting procedure;
in any case, as it will be evident from Fig.~\ref{F1}, the measured value 
of the cross section at 43.6 MeV/c can be hardly explained by a scattering 
length expansion of the annihilation cross section.  

%
\begin{table}[h!]
\centering
\caption{ $\bar{p}p$ total annihilation cross section at different
antiproton incident momenta multiplied by the velocity $\beta$ 
of the antiprotons [1,2]. In addition to the statistical and systematic error
an overall normalization error of 3.4\% has to be considered. 
}
\vskip 0.1 in                                                                   
\begin{tabular}{|c|c|c|} \hline                                  
     &  $\bar{p}$ incident & $\beta\sigma_{ann}^T$  \\
Ref. &  momentum           &   ($mbarn$)              \\
     &  ($MeV/c$)          &                          \\
\hline                                                                          
\hline                                                                          
              & $174.4\pm2.0$ & $40.5\pm0.5\;(stat)\;\pm0.5\;(sys)$\\
              & $106.6\pm4.5$ & $40.4\pm0.5\;(stat)\;\pm1.7\;(sys)$\\
              & $65.1\pm2.0$ & $43.1\pm0.7\;(stat)\;\pm2.5\;(sys)$ \\
              & $63.6\pm2.0$ & $43.6\pm0.7\;(stat)\;\pm2.6\;(sys)$ \\
\cite{OBELIX} & $62.1\pm2.2$ & $44.1\pm0.7\;(stat)\;\pm2.7\;(sys)$ \\
              & $60.5\pm2.2$ & $43.6\pm0.7\;(stat)\;\pm2.7\;(sys)$ \\
              & $54.4\pm2.8$ & $46.0\pm0.7\;(stat)\;\pm2.4\;(sys)$ \\  
              & $52.9\pm2.8$ & $46.4\pm0.7\;(stat)\;\pm2.5\;(sys)$ \\
              & $51.3\pm2.9$ & $47.0\pm0.8\;(stat)\;\pm2.7\;(sys)$ \\
              & $43.6\pm3.1$ & $55.2\pm0.9\;(stat)\;\pm4.1\;(sys)$ \\
\hline                                                                          
              & $54.9\pm2.5$ & $45.8\pm0.9\;(stat)\;\pm2.5\;(sys)$ \\
              & $53.8\pm2.6$ & $46.8\pm0.9\;(stat)\;\pm2.6\;(sys)$ \\
              & $52.6\pm2.6$ & $46.9\pm0.9\;(stat)\;\pm2.6\;(sys)$ \\
\cite{LEAP96} & $51.3\pm2.7$ & $47.0\pm0.9\;(stat)\;\pm2.6\;(sys)$ \\
              & $49.9\pm2.7$ & $47.1\pm0.9\;(stat)\;\pm2.7\;(sys)$ \\
              & $48.4\pm2.8$ & $47.5\pm0.9\;(stat)\;\pm2.7\;(sys)$ \\
              & $46.6\pm2.8$ & $48.4\pm0.9\;(stat)\;\pm2.7\;(sys)$ \\
\hline                                                                          
\end{tabular}
\label{TAB1}
\end{table}
In the momentum region considered and interesting here, from
47 to 106 MeV/$c$, only two partial waves contribute to the annihilation
cross section. Actually, as previously mentioned, we did not take into account
the difference between triplet and singlet contributions and worked
with spin averaged values of the scattering lengths. 
Consequently, in the theoretical expression of the cross section, we have 
four independent parameters corresponding to the real and imaginary parts 
of the S-wave scattering length and P-wave scattering volume.

%

In the previous section we emphasized that, in the range of momentum
considered, the cross section values are not much 
sensitive to the real part of the amplitudes. In particular, for the S-wave,
the contribution of the real part of the amplitude could be of the order of
few percents at the highest values of the momentum, whereas for the P-wave
it should be almost negligible. 
Since atomic data show that the absolute value of the real part is 
approximately equal to the one of the imaginary part \cite{Batty,CP92}, 
we choose the value of 0.85 fm for the real parts of the amplitudes,
corresponding to the averaged atomic ones.

Finally, we obtain two free parameters in the expression of the cross section:
the imaginary parts of the scattering amplitude and scattering volume. 
The fit was performed by means of the MINUIT program \cite{MINUIT} and 
provided the following best fit values for the two free parameters:
\begin{eqnarray*}
\mbox {Im} a_{sc} = - 0.69 \pm 0.01 (\mbox{stat}) \pm 0.03 (\mbox{sys}) \mbox { fm}; \\
\mbox {Im} A_{sc} = - 0.76 \pm 0.05 (\mbox{stat}) \pm 0.04 (\mbox{sys}) \mbox { fm}^3
\end{eqnarray*}
with the value of $\chi^2 = 0.2$ per point. In the fitting procedure only
the statistical errors were accounted for and produced the errors quoted
as statistical in the values of the best fit parameters. 
Concerning the systematic error of the data, the value 
shown in Tab.~\ref{TAB1} is dominated by the uncertainty in 
the beam momentum, which reflects on the $\beta\sigma$ values \cite{OBELIX};
if one assumes the values of the incident momentum at the center of the 
intervals, this systematic error does not affect the results of the fit.
However, the data are also affected by an overall normalization error of 3.4\%, 
essentially due to Monte Carlo corrections and beam normalization \cite{OBELIX}.
Changing all the experimental values of the cross section by 3.4\%, the 
two free parameters are modified within the systematic error quoted above.

In Fig.\ref{F1} the experimental and theoretical
values of the cross section multiplied by the square of the velocity 
$\beta$ of the antiproton
are shown; the errors quoted on the data are the statistical ones.  
The agreement of the 
theoretical predictions with the experimental data is excellent, 
apart from the point at 43.6 MeV/c, which is not considered in the fit and
appears in strong disagreement with the behaviour of the other data.   

These results obtained for the imaginary parts of the $\bar{p}p$ scattering 
length 
and scattering volume can be compared with the values of the same parameters 
determined using atomic data
(for the S-wave, we take an average value over a few experiments
\cite{Batty,CP92}; for the P-wave, the scattering volume is obtained 
from the average value of the 2P-level width by usual Trueman formula): 
\begin{eqnarray*}
\mbox {Im} a_{sc}  =  - 0.71 \pm 0.05 \mbox { fm}; \\
\mbox {Im} A_{sc} = - 0.71 \pm 0.07 \mbox { fm}^3.
\end{eqnarray*}
The agreement between the two sets of values is good. Nevertheless it is worth
making few comments, which could be important for the design of 
further experiments.

First, the information on the S-wave scattering length is obtained by using
the data on the shift and broadening of the ground state of protonium.
In the experiments made up to now, it was very difficult to control the
real weights of the singlet and triplet contributions. Therefore it is hardly
possible to say that the existing value of the scattering length, obtained from
atomic data, is an averaged value with usual statistical weights. However,
atomic experiments could be very informative for the determination of the
S-wave scattering lengths.
In fact, it would be possible to obtain separately singlet and triplet 
scattering
lengths if K$_{\alpha}$ X-rays will be measured in coincidence with
exclusive final states with definite initial state spin value.

Second, the information about P-wave scattering volume is indirect and is
obtained from the measurement of the total yield of L X-rays, together with
the yield of K$_{\alpha}$ X-rays. 
This procedure gives the spin averaged {\sl inverse}
value of the 2P-level width $\Gamma^{\rm inv}_{\rm 2P}$
\begin{eqnarray*}
\frac{12}{\Gamma^{\rm inv}_{\rm 2P}} = 
\frac{1}{\Gamma(^3P_0)} +\frac{3}{\Gamma(^1P_1)} +\frac{3}{\Gamma(^3P_1)} +
\frac{5}{\Gamma(^3P_2)}.
\end{eqnarray*}
This formula assigns ``inverse" weights to different partial waves: 
the larger is
the value of a partial wave, the smaller is its contribution. Such a procedure
can be effective if one wants to obtain an approximative value of the 
2P-level width and P-wave scattering volume, 
however it can not be confidently used to determine the correct spin averaged
value of $\Gamma_{\rm 2P}$.
 
It has been stressed that, at these low energies, the effect of the Coulomb 
interaction must necessarily accounted for in the expression of 
the cross section.  The comparison of low energy antiproton-proton
annihilation cross section behaviour with and without Coulomb interaction
\cite{CP93,LEAP96-CP} shows that the difference between these cross sections
is very large, even at the present LEAR energies. For instance, for
antiproton laboratory momentum of 60 MeV/$c$, this difference is of the
order of 100 mb.   
 
Nevertheless, it may be interesting and instructive to perform the fit 
of the low energy cross section using the wrong
expression of the amplitude, which does not take into account 
the Coulomb interaction. The fit with the wrong formula still interpolates the
experimental
points, however the agreement is worse than in the previous case
($\chi^2=0.9$ per point) and the imaginary part of the S-wave scattering
length is set to very big value of $-1.13\pm0.01$ fm which is incompatible 
with other experimental data.

\section{Conclusions}

The analysis of the recent experimental data obtained by the OBELIX experiment 
shows that the low energy parameters of the $\bar{p}p$ scattering 
amplitude (the imaginary parts of the S-wave scattering
length and P-wave scattering volume), extracted from the data on
antiproton-proton annihilation at low energy, are in very good agreement with
existing atomic data and are of quite good precision.

For further improvements, it would be very useful to measure the low energy
annihilation cross section with higher accuracy and at still lower energies. 
This kind of
experiments could be done with the new facility AD (Antiproton Decelerator),
which is under development. The measurement of the cross section at low
energy would be the most effective for the determination of the 
P-wave scattering volume, whereas the atomic data
would be preferable to extract S-wave scattering lengths.  

Finally, it would be also very instructive to measure, at very low energy,
those particular channels of
the $\bar{p}p$ annihilation for which the annihilation in
S-state is forbidden by selection rules (for instance, 
$\eta \eta$ or $\pi^0\pi^0$).
Nevertheless, due to Coulomb effects, the annihilation cross section
will be different from zero, since the P-wave annihilation does 
not vanishing at low energies. 
It would be, probably, the first measurement 
of reaction cross sections near threshold ``forbidden''
without Coulomb interaction and ``allowed'' in the presence of it.   

Finally, another possibility to obtain the same information is offered
by experiments in the traps \cite{CP96}, where, for instance, the measurement of
the time of life of antiprotons in a trap allows the
imaginary part of scattering amplitudes to be extracted as well.

\end{document}